# Nearly Perfect Spin Polarization of Noncollinear Antiferromagnets

Gautam Gurung[1,2,3,*], Mohamed Elekhtiar[1], Qing-Qing Luo[4,5], Ding-Fu Shao[4,†], and Evgeny Y. Tsymbal[1,‡]

[1] *Department of Physics and Astronomy & Nebraska Center for Materials and Nanoscience, University of Nebraska, Lincoln, Nebraska 68588-0299, USA*

[2] *Clarendon Laboratory, Department of Physics, University of Oxford, Parks Road, Oxford, OX1 3PU UK*

[3] *Trinity College, University of Oxford, Oxford, OX1 3BH UK*

[4] *Key Laboratory of Materials Physics, Institute of Solid-State Physics, HFIPS, Chinese Academy of Sciences, Hefei 230031, China*

[5] *University of Science and Technology of China, Hefei 230026, China*

Ferromagnets with high spin polarization are known to be valuable for spintronics—a research field that exploits the spin degree of freedom in information technologies. Recently, antiferromagnets have emerged as promising alternative materials for spintronics due to their stability against magnetic perturbations, absence of stray fields, and ultrafast dynamics. For antiferromagnets, however, the concept of spin polarization and its relevance to the measured electrical response are elusive due to nominally zero net magnetization. Here, we define an effective momentum-dependent spin polarization and reveal an unexpected property of many noncollinear antiferromagnets to exhibit nearly 100% spin polarization in a broad area of the Fermi surface. This property leads to the emergence of an extraordinary tunneling magnetoresistance (ETMR) effect in antiferromagnetic tunnel junctions (AFMTJs). As a representative example, we predict that a noncollinear antiferromagnet $Mn_3GaN$ exhibits nearly 100% spin-polarized states that can efficiently tunnel through low-decay-rate evanescent states of perovskite oxide $SrTiO_3$ resulting in ETMR as large as $10^4$%. Our results uncover hidden functionality of material systems with noncollinear spin textures and open new perspectives for spintronics.

## Introduction

Materials with high spin polarization have been of significant interest for applications in spintronics—a research field that exploits the spin degree of freedom for information technologies[1]. Qualitatively, spin polarization can be understood as the extent to which the spin of electrons is aligned with a certain direction. Quantitatively, however, the spin polarization is not uniquely defined and can be referred either to the uneven number of up-spin and down-spin electrons at the Fermi energy or to the unbalanced (spin-polarized) currents carried by electrons with opposite spin orientations[2]. Even in the latter case, the transport spin polarization appears to be different as determined from spin-dependent tunneling[3] or ballistic transmission[4,5] experiments. Nonetheless, whatever the definition is used, a high degree of spin polarization, ideally 100%, is beneficial for spintronics. This is due to a stronger electric response that can be achieved in transport measurements. For example, half-metals—ferromagnetic materials that have only one spin state at the Fermi energy and thus 100% spin polarization[6] (independent of the definition)—are supposed to exhibit an infinitely large tunneling magnetoresistance (TMR) if used as electrodes in magnetic tunnel junctions (MTJs)[7].

The recent interest and progress in antiferromagnetic (AFM) spintronics[8,9] puts forward the spin degree of freedom in AFM metals and the AFM Néel vector as a state variable. For AFM metals, however, the concept of spin polarization and its relevance to the measured electric response becomes even more subtle. Most antiferromagnets host $\hat{P}\hat{T}$ and/or $\hat{T}\hat{t}$ symmetries, where $\hat{P}$ is space inversion, $\hat{T}$ is time reversal, and $\hat{t}$ is half a unit cell translation, which make their band structures spin-degenerate and thus spin polarization vanishing. There exists however a class of antiferromagnets that have broken $\hat{P}\hat{T}$ and $\hat{T}\hat{t}$ symmetries and thus exhibit a spin-split band structure, including certain types of noncollinear[10,11] and collinear[12-15] antiferromagnets. Due to the alternating local crystallographic environment along the two magnetic sublattices the latter were dubbed altermagnets[16]. Recently, it has been proposed that the concept of altermagnetism can be extended to accommodate noncollinear spins and multiple local structure variations[17].

In antiferromagnets with violated $\hat{P}\hat{T}$ and $\hat{T}\hat{t}$ symmetries, Kramers' spin degeneracy is broken even in the absence of spin-orbit coupling, and hence these materials can support longitudinal spin-polarized currents along certain crystallographic orientations[11,18,19]. Due to this property, while the net magnetization of these antiferromagnets is zero, their transport spin polarization is not, which allows using them as ferromagnets in spintronic devices. It is not obvious, however, how the net transport spin polarization of the antiferromagnets is related to their resistive response of a spintronic device such as, e.g., an AFM tunnel junction (AFMTJ)[20-25]. In fact, it has been predicted that a large TMR in AFMTJs can occur even if the net currents are spin-neutral, i.e., the net transport spin polarization is zero[20]. The TMR effect in these junctions relies on the conservation of the transverse momentum in the process of tunneling and controlled by matching the spin-polarized Fermi surfaces of the two AFM electrodes. As a result, the net spin polarization of the electrodes is less relevant, while the spin polarization of conduction channels plays an essential role[20].

In this work, we define an effective momentum-dependent spin polarization and reveal an unexpected property of many noncollinear antiferromagnets to exhibit nearly perfect spin polarization in a broad area of the Fermi surface. As a result, using these non-collinear antiferromagnets in AFMTJs leads to an extremely high (extraordinary) TMR effect.

## Results

**Spin Polarization.** Here, conduction channels are defined as propagating Bloch states in a metal electrode which are determined by transverse wave vector $k_\parallel$, band number $n$, and spin $s_{nk_\parallel}$. Since $k_\parallel$ is conserved in the tunneling process, it is the spin state of conduction channels in the electrodes that controls the TMR effect in AFMTJs.

The spin state can be quantified in terms of the net spin $s_{k_\parallel}$ of conduction channels at the transverse wave vector $k_\parallel$:

$$s_{k_\parallel} = \sum_n s_{nk_\parallel} = \sum_n \frac{l_z}{2\pi} \int \langle \psi_{nk} | s | \psi_{nk} \rangle \delta(E_{nk} - E_F) dk_z. \quad (1)$$

Here $l_z$ is the lattice constant of the electrode along the transport $z$ direction, $\langle \psi_{nk} | s | \psi_{nk} \rangle$ is the spin expectation value for band $n$ of energy $E_{nk}$ and eigenfunction $\psi_{nk}$ at wave vector $k = (k_\parallel, k_z)$, and $E_F$ is the Fermi energy. For collinear AFM metals, spin $s_{k_\parallel}$ is allowed to have only two directions: up, i.e., parallel to the Néel vector (that is parallel to one of the AFM sublattices), or down, i.e., antiparallel to the Néel vector. Its magnitude, $s_{k_\parallel} \equiv |s_{k_\parallel}| = N_{k_\parallel}^\uparrow - N_{k_\parallel}^\downarrow$, is determined by the number of conduction channels $N_{k_\parallel}^{\uparrow,\downarrow}$ for up- ($\uparrow$) and down- ($\downarrow$) spin electrons at $k_\parallel$. In this case, the net spin determines the spin polarization of conduction channels given by $p_{k_\parallel} = \frac{s_{k_\parallel}}{\sum_n |s_{nk_\parallel}|} = \frac{N_{k_\parallel}^\uparrow - N_{k_\parallel}^\downarrow}{N_{k_\parallel}}$, where $N_{k_\parallel}$ is the total number of conduction channels at $k_\parallel$.

For collinear sublattice magnetizations, in the absence of spin-orbit coupling, spin is conserved in the tunneling process, and therefore transmission at $k_\parallel$, $T(k_\parallel)$, is controlled by matching the spin components of the wave functions in the electrodes. For example, for an AFMTJ with parallel (P) Néel vectors of the two AFM electrodes, $T(k_\parallel)$ is large, while for antiparallel (AP) Néel vectors, $T(k_\parallel)$ is small due to a mismatch of the spin states in the AFM electrodes. As a result, the total transmission for the P state ($T_P$) is larger than that for the AP state ($T_{AP}$), leading to a non-zero TMR ratio, $TMR = (T_P - T_{AP})/T_{AP}$.

On the contrary, for noncollinear AFM metals, spin is not a good quantum number and, in general, is not conserved in the tunneling process (even in the absence of spin-orbit coupling). The spin magnitude and direction vary depending on $k_\parallel$ and $n$. Nevertheless, one can define an *effective* spin polarization,

$$p_{k_\parallel} = \frac{s_{k_\parallel}}{\sum_n |s_{nk_\parallel}|}, \quad (2)$$

as a vector that has different magnitudes and orientations at different $k_\parallel$. Due to being $k_\parallel$-dependent, the effective spin polarization is different from the conventional spin polarization which is defined with respect to a global spin-quantization axis. As seen from the definition, the spin polarization magnitude, $p_{\mathbf{k}_\parallel} \equiv |\mathbf{p}_{\mathbf{k}_\parallel}|$, is equal to 100% when spins $\mathbf{s}_{\mathbf{k}_\parallel n}$ are parallel in all conduction channels $n$ at $\mathbf{k}_\parallel$ or only one conduction channel is present. This situation is reminiscent of the collinear case, where spin is conserved in the process of tunneling. As a result, matching the net spins $\mathbf{s}_{\mathbf{k}_\parallel}$ in the electrodes can be used as the necessary requirement for large transmission $T(\mathbf{k}_\parallel)$ in the areas of the Brillouin zone where $p_{\mathbf{k}_\parallel}$ is close to 100%. On the contrary, at those $\mathbf{k}_\parallel$ where there are a few conduction channels with non-collinear $\mathbf{s}_{n\mathbf{k}_\parallel}$, spin is not conserved and thus $\mathbf{p}_{\mathbf{k}_\parallel}$ could not serve as a proper measure of TMR in the spirit of Julliere's formula[7].

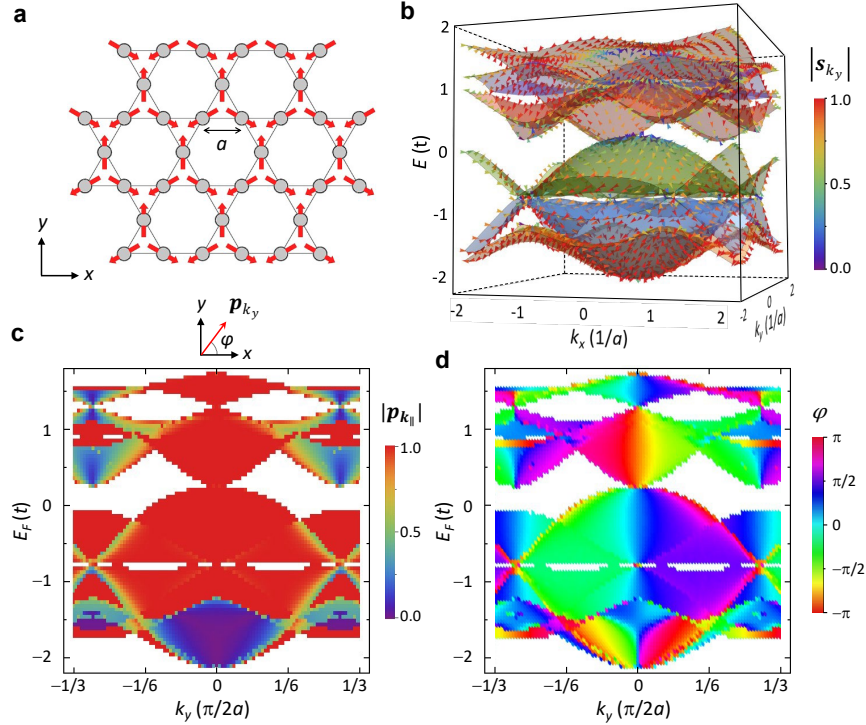

**Fig. 1 Spin polarization of a noncollinear Kagome antiferromagnet.** (**a**) Kagome lattice with non-collinear AFM structure of magnetic moments. (**b**) Calculated band structure of the Kagome lattice for $\Delta/t = 1.5$. Arrows indicate the spin expectation values. Spin magnitudes are shown in color. (**c,d**) Map of the spin polarization vector $p_{k_y}$ as a function of $k_y$ and $E_F$ for the Kagome lattice (**a**) for $\Delta/t = 1.5$. $p_{k_y}$ vectors lie in the $x$-$y$ plane ($p_z = 0$). Their magnitudes $p_k \equiv |p_k|$ (**c**) and polar angles $\varphi$ (**d**) are depicted in color.

An important implication following from this observation, is the possibility of having a 100% spin polarization in non-collinear antiferromagnets. To illustrate this property, we consider a simple tight-binding model of a Kagome lattice with magnetic moments forming a noncollinear AFM configuration, as shown in Figure 1a. This magnetic structure mimics a noncollinear two-dimensional (2D) antiferromagnet with broken $\hat{P}\hat{T}$ and $\hat{T}\hat{t}$ symmetries. Assuming one orbital per atom with exchange-split on-site energies $E_i^{\uparrow,\downarrow} = E_i \pm \frac{\Delta}{2}$ ($i = 1,2,3$) and spin-independent nearest-neighbor hopping $t$, we arrive at the band structure $E(\mathbf{k})$ ($\mathbf{k} = k_x, k_y$) that consists of six bands with $\mathbf{k}$-dependent spin expectation values, as shown in Figure 1b (see Methods for details of the calculation). Figures 1c,d show magnitudes $p_k \equiv |\mathbf{p}_k|$ and polar angles $\varphi$ of the calculated transport spin polarization $\mathbf{p}_{k_y}$ (here we assume $x$-direction for transport) as a function of transverse wave vector $k_y$ and the Fermi energy $E_F$. It is seen that while $\mathbf{p}_{k_y}$ is oriented in different directions (Fig. 1d), its magnitude is nearly 100% (indicated by red color in Fig. 1c) in a broad range of $E_F$. Only at small $E_F$, $|\mathbf{p}_{k_y}|$ is significantly reduced due to the overlap of bands with different spin orientations (Fig. 1c). Qualitatively, the large spin polarization is sustained in a broad range of $k_y$ and $E_F$ and only at small values of the spin splitting $\Delta$ it gets reduced (Supplementary Fig. S1[26]).

**Extraordinary TMR.** In three-dimensional (3D) spin-split antiferromagnets, the Fermi surface can form regions with 100% spin polarization in a two-dimensional (2D) Brillouin zone (2DBZ) perpendicular to the transport direction. These fully spin-polarized Bloch states can generate a very large TMR effect in the associated AFMTJs (Fig. 2e). The mechanism of such TMR is schematically illustrated in Figure 2f showing conduction between left and right AFM electrodes for parallel (P) (top) and antiparallel (AP) (bottom) Néel vectors in $k$-space. The electrodes have fully spin-polarized conduction channels indicated in Figure 2f by two crossing ellipses with the spin polarization vector having different orientation $\theta$ (indicated by varying color). If electronic states in these channels dominate transmission, TMR in such AFMTJ is expected to become virtually infinite. This is due to the spin-state match of the 100% spin-polarized conduction channels in the left and right electrodes for parallel-aligned AFMTJ and the spin-state mismatch for antiparallel. In contrast to the ordinary TMR in conventional MTJs with ferromagnetic (FM) electrodes (Fig. 2a,b) and in AFMTJs with collinear exchange-split AFM (altermagnetic) electrodes (Figs. 2c,d), where the spin polarization is defined with respect to the global quantization axis, the predicted TMR effect (Figs. 2e,f) relies on of the momentum-dependent spin polarization of a non-collinear antiferromagnet, which may have different orientations at different $k_\parallel$ (Eq. 2). While such TMR seems qualitatively similar to the ultimately infinite TMR in MTJs with ideal half-metallic electrodes, there is a conceptual difference between the two effects. In nominal half metals, the fully spin-polarized conduction arises from the presence of only one-spin bands at the Fermi energy, resulting in the global 100% spin polarization of the electric current, whereas in non-collinear antiferromagnets, the fully spin-polarized conduction occurs only within a given $k_\parallel$-defined conduction channel with the global spin polarization of electric current being incomplete (or even possibly zero). Therefore, since the mechanism of the giant TMR effect in non-collinear AFMTJs is conceptually different from the ordinary TMR in collinear MTJs and AFMTJs, we dub it extraordinary TMR (ETMR).

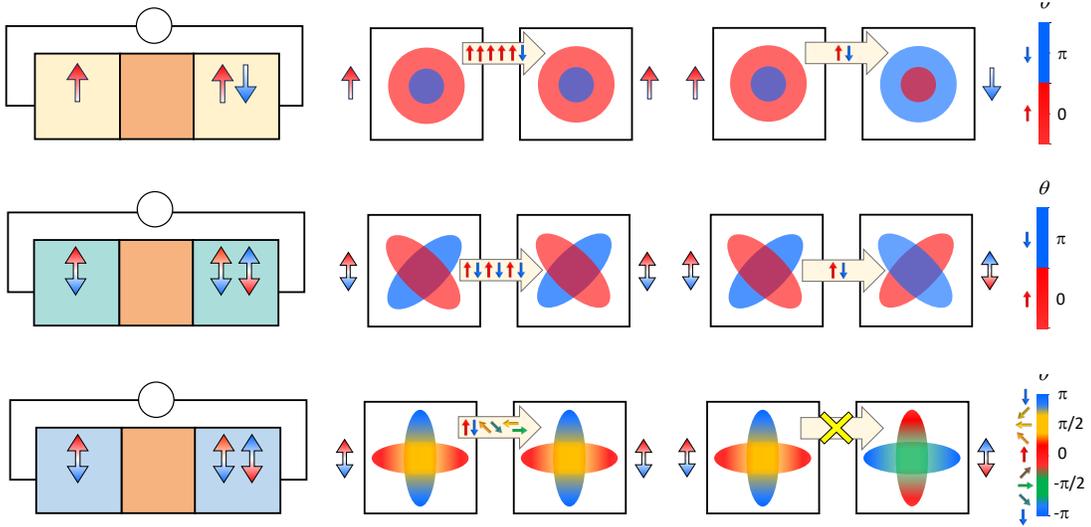

**Fig. 2 Schematics of different types of tunnel junctions and TMR effects.** (**a**) Schematics of a conventional MTJ where two ferromagnetic (FM) electrodes are separated by a tunnel barrier. (**b**) Mechanism of TMR in a conventional MTJ. Conduction (indicated by block arrows) between left and right FM electrodes for parallel (P) and antiparallel (AP) magnetization (indicated by large arrows). Red and blue circles denote the Fermi surfaces of the ferromagnet for up-spin (small red arrows) and down-spin (small blue arrows) electrons. Angle $\theta$ refers to the spin orientation with respect to the magnetization in the left electrode. (**c**) Schematics of an AFMTJ with collinear exchange-split AFM electrodes (C-AFMTJ). (**d**) Mechanism of TMR in C-AFMTJ. Conduction between left and right AFM electrodes for P and AP Néel vectors (indicated by double-arrows). Red and blue ellipses denote the Fermi surfaces of the antiferromagnet for up- and down-spin electrons. Angle $\theta$ refers to the spin orientation with respect to the Néel vector in the left electrode. (**e**) Schematics of an AFMTJ with non-collinear AFM electrodes (NC-AFMTJ). (**f**) Mechanism of ETMR. The NC-AFM electrodes have fully spin-polarized conduction channels within the area indicated by crossing ellipses with the spin polarization vector having different orientation $\theta$ (indicated by varying color). Electrons in these channels can efficiently tunnel through a tunnel barrier due to its low-decay-rate evanescent states supporting transmission. Matching (mismatching) of the 100% polarized conduction channels in the two electrodes for the P (AP) state produces ETMR.

To provide the required efficient tunneling of the fully spin-polarized states, low-decay-rate evanescent states in the tunneling barrier need to support their transmission. Therefore, for observing ETMR, in addition to 100% spin-polarized conduction channels in the AFM electrodes, their distribution in the 2DBZ needs to be matched to the distribution of the low-decay-rate evanescent states in the insulator at the Fermi energy. Thus, we identify two important properties which may exhibit non-collinear AFM metals and the associated AFMTJs: nearly 100% transport spin polarization and ETMR.

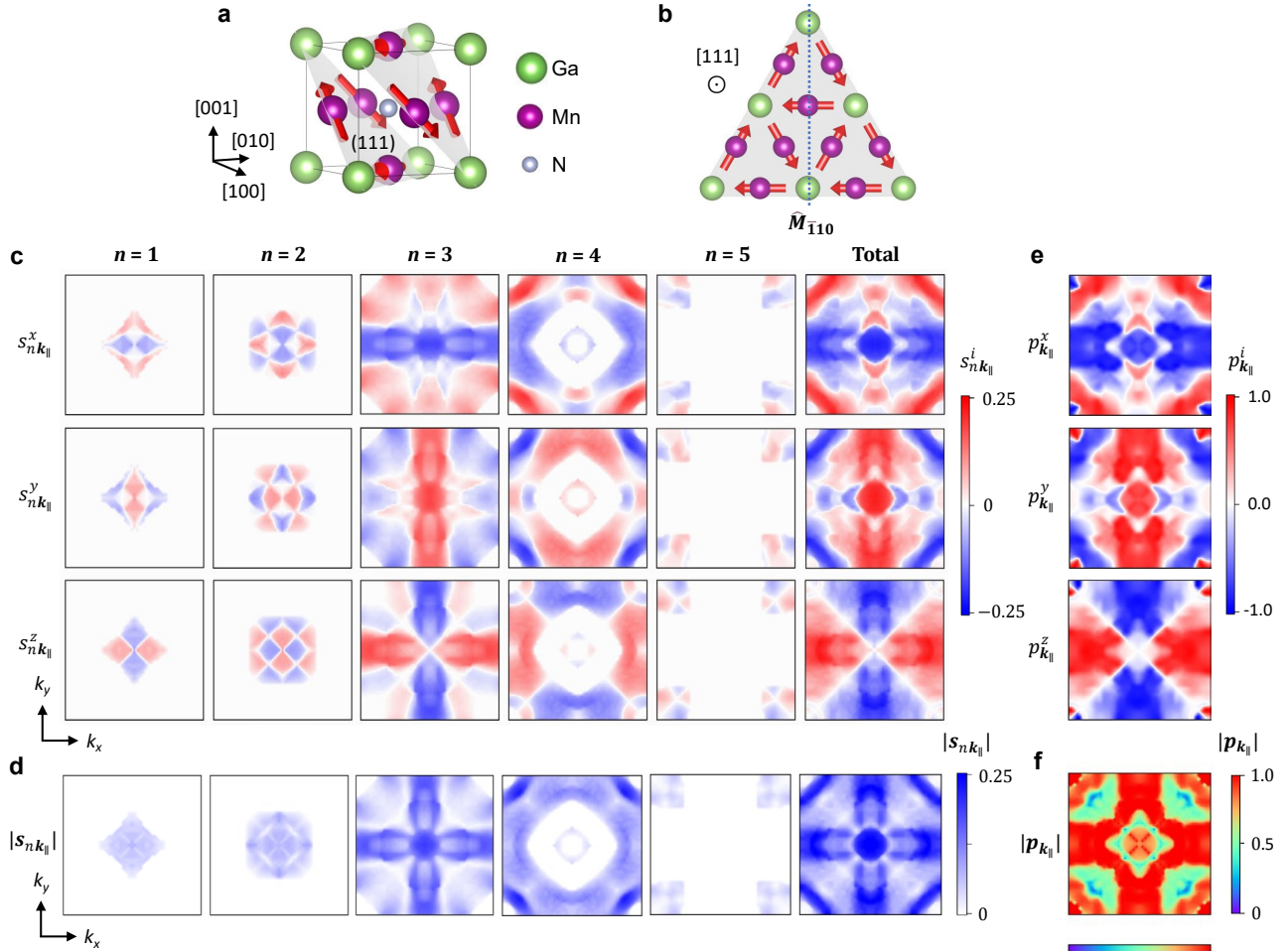

**Fig. 3 Atomic structure, spin texture and effective spin polarization at the Fermi surface of Mn$_3$GaN.** (**a,b**) Atomic and magnetic structure of antiperovskite Mn$_3$GaN in the noncollinear $\Gamma_{5g}$ AFM configuration in 3D view (**a**) and projected to the (111) plane with $\widehat{M}_{\bar{1}10}$ and $\hat{C}_{\bar{1}10}$ symmetries indicated (**b**). (**c, d**) Components of spin $s_{nk_\parallel} = (s^x_{nk_\parallel}, s^y_{nk_\parallel}, s^z_{nk_\parallel})$ (c) and spin magnitudes $s_{nk_\parallel} \equiv |s_{nk_\parallel}|$ (**d**) for five bands (labeled by index $n$) contributing to the Fermi surface of Mn$_3$GaN and plotted in the 2DBZ of Mn$_3$GaN (001). (**e,f**) Components of the effective spin polarization $p_{k_\parallel} = (p^x_{k_\parallel}, p^y_{k_\parallel}, p^z_{k_\parallel})$ (**e**) and polarization magnitude $p_{k_\parallel} \equiv |p_{k_\parallel}|$ (**f**).

**Spin polarization of Mn$_3$GaN (001).** In the following, we demonstrate that these properties can be observed in practice. As a representative example of a noncollinear antiferromagnet, we consider antiperovskite Mn$_3$GaN, and show, based on density-functional theory (DFT) calculations (see Methods for details), that this antiferromagnet hosts conduction channels with nearly 100% effective spin polarization in a broad area of the Fermi surface. We then show that these highly spin-polarized states in antiperovskite Mn$_3$GaN match the evanescent states with low decay rates in perovskite oxide SrTiO$_3$ resulting in ETMR as large as $10^4$% in Mn$_3$GaN/SrTiO$_3$/Mn$_3$GaN (001) AFMTJs.

Mn$_3$XN-type (X = Ga, Sn, Ni …) antiperovskite crystals have a cubic structure similar to perovskites, except the positions of anions and cations being interchanged. The frustrated Mn-kagome lattice in the (111) plane favors a noncollinear AFM ordering, resulting in interesting spin-dependent properties [27-31]. Figure 3a shows the atomic and magnetic structure of Mn$_3$GaN, where Mn magnetic moments form a 120° chiral configuration within the (111) plane (Fig. 3b). Such a $\Gamma_{5g}$ noncollinear AFM structure breaks $\hat{P}\hat{T}$ and $\hat{T}\hat{t}$ symmetries, resulting in a spin-polarized Fermi surface (Supplementary Fig. S2[26]) and hence spin-polarized conduction channels.

Figure 3c shows the calculated spin components $s_{n\bm{k}_\parallel} = (s^x_{n\bm{k}_\parallel}, s^y_{n\bm{k}_\parallel}, s^z_{n\bm{k}_\parallel})$ for each of five bands (labeled by index $n$) contributing to the Fermi surface of Mn$_3$GaN (001). While all bands are spin textured, the largest contribution to the total spin $s_{\bm{k}_\parallel} = \sum_n s_{n\bm{k}_\parallel}$ (Fig. 3c, rightmost panels) comes from bands 3 and 4 that have the largest Fermi surfaces (Fig. 3c, panels labeled by $n = 3$ and $n = 4$). Band 3 has a pronounced cross feature in its spin texture, while band 4 reveals the largest spin values along the diagonal lines close to the corners of the 2DBZ (Fig. 3d). It is notable that for all bands, $s^z_{\bm{k}_\parallel}$ vanishes along the diagonals, [110] and [$\bar{1}$10], in the 2DBZ (Fig. 3c, third row). This is due to mirror symmetry $\hat{M}_{\bar{1}10}$ and two-fold rotation symmetry $\hat{C}_{\bar{1}10}$ (Fig. 3b) supporting transformations of the spin component $s^z_{\bm{k}_\parallel}$ at wave vector $\bm{k}_\parallel = (k_x, k_y)$ as follow: $\hat{M}_{\bar{1}10} s^z_{k_x k_y} = -s^z_{k_y k_x}$ and $\hat{C}_{\bar{1}10} s^z_{k_x k_y} = -s^z_{-k_y -k_x}$. As a result, $s^z_{\bm{k}_\parallel}$ is antisymmetric with respect to diagonals [110] and [$\bar{1}$10] of the 2DBZ and zero at the diagonals (see also Supplementary Fig. S3[26]).

All these features lead to the net spin texture with the largest $s_{\bm{k}_\parallel} \equiv |\bm{s}_{\bm{k}_\parallel}|$ appearing around the $k_x = 0$ and $k_y = 0$ lines in the 2DBZ, forming a cross pattern, and at the diagonal lines close to the corners of the 2DBZ (Fig. 3d, rightmost panel). This spin texture is mirrored by the momentum-dependent spin polarization $\bm{p}_{\bm{k}_\parallel}$. As seen from Figure 3e, the $p^i_{\bm{k}_\parallel}$ ($i = x, y, z$) components are reminiscent to the $s^i_{\bm{k}_\parallel}$ components (Fig. 3c,

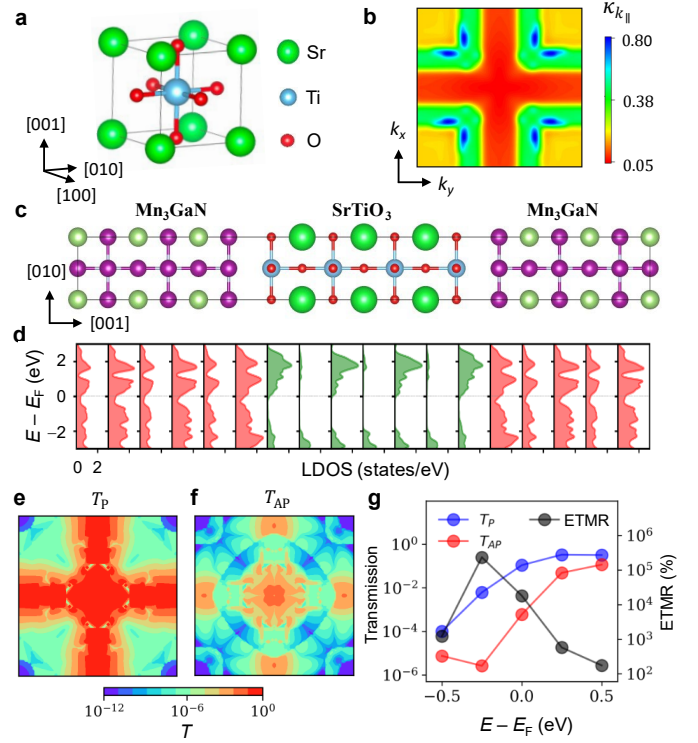

**Fig. 4 ETMR effect in Mn$_3$GaN/SrTiO$_3$/Mn$_3$GaN (001) AFMTJ** (**a**) Atomic structure of perovskite oxide SrTiO$_3$. (**b**) The lowest decay rate $\kappa_{\bm{k}_\parallel}$ of the evanescent states of SrTiO$_3$ (001) in the 2DBZ calculated at the energy close to the bottom of the conduction band. (**c**) Atomic structure of the AFMTJ. The Néel vectors of the Mn$_3$GaN electrodes lie in the [111] plane (not shown) and can be parallel or antiparallel. (**d**) Layer-resolved density of states (LDOS) for parallel Néel vectors. (**e,f**) Calculated $\bm{k}_\parallel$-resolved transmission in the 2DBZ for the AFMTJ for parallel (P) (**e**) and antiparallel (AP) (**f**) alignment of the Néel vector in Mn$_3$GaN electrodes. (**g**) Total transmission and ETMR ratio as functions of energy.

rightmost panels), and the cross pattern featuring the distribution of $s_{\bm{k}_\parallel}$ (Fig. 3d, rightmost panel) is mimicked by the distribution of $p_{\bm{k}_\parallel} \equiv |\bm{p}_{\bm{k}_\parallel}|$ in Figure 3f.

The most important observation following from these results is nearly 100% spin polarization in a broad area of the 2DBZ of Mn$_3$GaN (001), especially around the $k_x = 0$ and $k_y = 0$ lines away from the zone center (Fig. 3e). This feature emerges due to bands 3 and 4 having nearly the same spin orientation around these lines and no other bands appearing in these regions (Fig. 3c). The somewhat reduced spin polarization near the zone center is caused by overlap of the non-collinear spin states in bands 1, 2, 3, and 4. Interestingly, we observe that some regions in the 2DBZ of Mn$_3$GaN (001) exhibit a persistent spin texture, where spins in relatively broad region of the 2DBZ are pointing in the same direction (Supplementary Fig. S4). We note however that this persistent spin texture is not symmetry enforced like that predicted in Ref. 32, but results from specific type of interactions intrinsic to a particular non-collinear antiferromagnet.

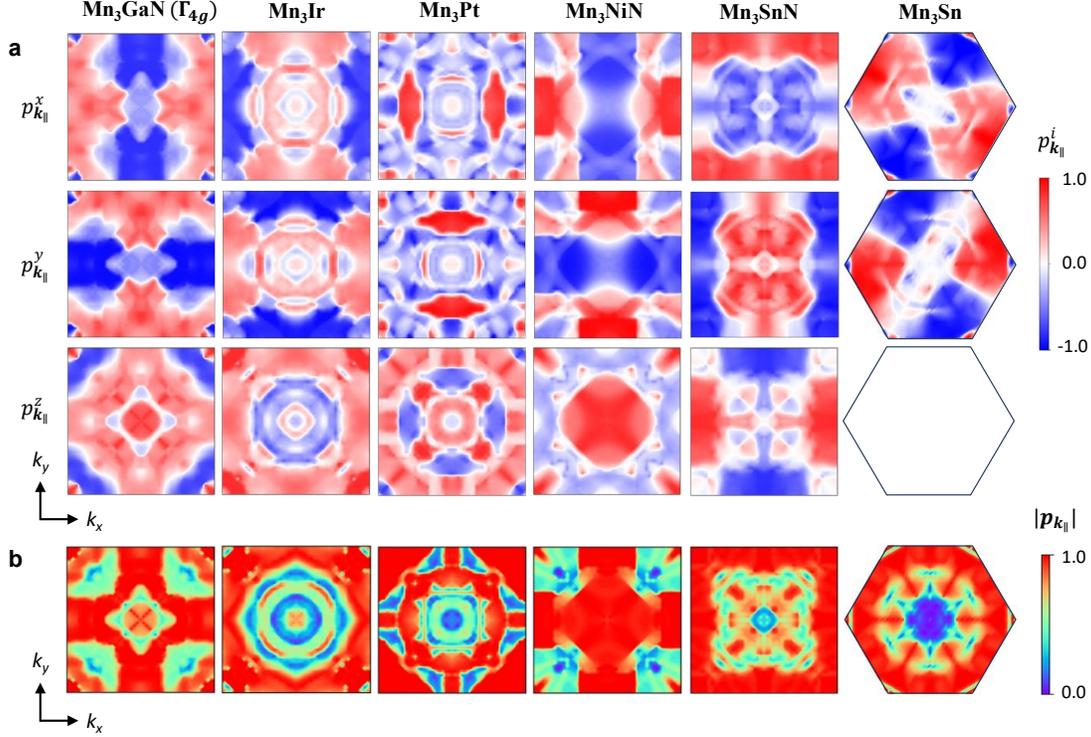

**Fig. 5 Effective spin polarization of different noncollinear antiferromagnets.** (a,b) Calculated components of spin polarization $p_{k_\parallel} = (p^x_{k_\parallel}, p^y_{k_\parallel}, p^z_{k_\parallel})$ (a) and spin polarization magnitudes $p_{k_\parallel} \equiv |p_{k_\parallel}|$ (b) for Mn$_3$GaN in the $\Gamma_{4g}$ phase, Mn$_3$Ir ($\Gamma_{4g}$), Mn$_3$Pt ($\Gamma_{4g}$), Mn$_3$NiN ($\Gamma_{4g}$), Mn$_3$SnN ($\Gamma_{5g}$), and Mn$_3$Sn.

**ETMR in Mn$_3$GaN/SrTiO$_3$/Mn$_3$GaN (001) AFMTJs.** As was discussed above, the appearance of 100% spin polarized states in a non-collinear AFM metal can be detected via the ETMR effect using an AFMTJ that supports efficient tunneling of these states. This requires an insulator whose complex band structure exhibits the distribution of the lowest-decay-rate evanescent states in the 2DBZ similar to the distribution of the 100% spin-polarized conduction channels in the AFM electrode. It appears that perovskite SrTiO$_3$ (Fig. 4a) satisfies this condition[33]. Figure 4b shows the lowest decay rate $\kappa_{k_\parallel}$ of the evanescent states in SrTiO$_3$ (001) in the 2DBZ (see Methods for details of these calculations). It is evident that the distribution of $\kappa_{k_\parallel}$ exhibits a cross pattern alike the distribution of the spin polarization $p_{k_\parallel}$ in Mn$_3$GaN (001) (Fig. 3f). Such matching between the high-$p_{k_\parallel}$ and low-$\kappa_{k_\parallel}$ areas in the 2DBZ suggests that a Mn$_3$GaN/SrTiO$_3$/Mn$_3$GaN (001) AFMTJ can be employed to detect the perfectly spin-polarized states in Mn$_3$GaN by measuring the ETMR effect.

Due to similar atomic structures and lattice constants, high-quality crystalline Mn$_3$GaN/SrTiO$_3$/Mn$_3$GaN (001) AFMTJs are feasible in practice. In fact, epitaxial Mn$_3$GaN films have been grown on SrTiO$_3$ revealing high crystallinity of their interface structure[34]. Figure 4c shows the atomic structure of the AFMTJ that is used in our calculations. Here a 3-unit-cell SrTiO$_3$ (001) barrier layer is placed between 2.5-unit-cell Mn$_3$GaN layers. The layers are connected across Mn$_2$N/TiO$_2$ interfaces which have the lowest energy among other interfaces[35]. We find that a wide band gap of SrTiO$_3$ is well maintained across the junction (Fig. 4d). The Mn$_3$GaN/SrTiO$_3$/Mn$_3$GaN (001) structure in Figure 4c is then used as the scattering region of the AFMTJ connected to two semi-infinite Mn$_3$GaN (001) electrodes for transmission calculations (see Methods for details).

We find for the P state, where the Néel vectors of the two electrodes are parallel, $T_P(k_\parallel)$ is strongly enhanced in a cross-pattern area of the 2DBZ, resulting from the high-$p_{k_\parallel}$-low-$\kappa_{k_\parallel}$ matching (Fig. 4e). We note that, as follows from Supplementary Section E, the interface structure of the AFMTJ maintains bulk-like features of the spectral density (Supplementary Fig. S5a) and its spin polarization (Supplementary Fig. S5b) in Mn$_3$GaN and the decay rate in SrTiO$_3$ (Supplementary Fig. S5c). In contrast, for the AP state, where the Néel vectors of the two electrodes are antiparallel, while the largest $T_{AP}(k_\parallel)$ also appears at the cross-pattern area, it is significantly reduced compared to $T_P(k_\parallel)$ due to reversed $s_{k_\parallel}$ in the two electrodes (Fig. 4f). As a result, the total transmission $T_{AP}$ is much smaller than $T_P$, producing TMR as large as $\sim 1.8 \times 10^4$% (Fig. 4g). This TMR value is gigantic, significantly larger than the values known for conventional MTJs and reminiscent to an infinitely large TMR expected for MTJs based on ideal half-metallic electrodes. In fact, estimating the spin polarization $P$ of the electrodes with Julliere's formula[7] $TMR = \frac{2P^2}{1-P^2}$, we obtain $P \approx 99.99\%$. This extraordinary behavior of the Mn$_3$GaN/SrTiO$_3$/Mn$_3$GaN (001) AFMTJ is due to the property of antiperovskite Mn$_3$GaN to exhibit fully spin-polarized electronic states that can efficiently tunnel through perovskite SrTiO$_3$ while preserving their spin state—the signature of the ETMR effect.

We note that in our calculation, the Fermi energy ($E_F$) of the AFMTJ lies near the conduction band minimum (CBM) of SrTiO$_3$ (Fig. 4d), while it is expected to appear well within the band gap of the insulator. This is due to the underestimated band gap of SrTiO$_3$ resulting from the well-known deficiency of DFT to correctly describe the excited states. Such a shortage, however, does not affect our main conclusions, since the ETMR appears not only for $E = E_F$ but also in a broad energy window around $E_F$ (Fig. 4g). Especially, we obtain even larger ETMR value of $\sim 2.3 \times 10^5\%$ at $E = E_F - 0.25$ eV, well inside the band gap of SrTiO$_3$, indicating the validity of our results. This enhancement of the ETMR ratio at $E = E_F - 0.25$ eV followed by some reduction at $E = E_F - 0.5$ eV (Fig. 4g) clearly correlates with the appearance of 100% spin polarization in the 2DBZ of Mn$_3$GaN at different energies (Supplementary Fig. S6). In addition, we find that the cross feature of the evanescent states persists much deeper in the band gap of SrTiO$_3$ (Supplementary Fig. S7[26]), suggesting that independent of the band offset between Mn$_3$GaN and SrTiO$_3$, matching between the highly polarized conducting channels in Mn$_3$GaN and low decay rate evanescent states in SrTiO$_3$ is well maintained. Note that the TMR ratio is reduced for energies above the CBM of SrTiO$_3$, which is expected since at the energies within the conduction band of SrTiO$_3$, the tunneling mechanism of conduction breaks down (Fig. 4g).

We also note that our calculations neglect the effects of spin-orbit coupling which may lead to spin mixing affecting the spin polarization and ETMR. We find, however, that taking spin-orbit coupling into account does not change our main conclusions. This is evident from Supplementary Section H, which shows the effects of spin-orbit coupling on the spin polarization of bulk Mn$_3$GaN (Supplementary Figs. S8a,b) and ETMR in Mn$_3$GaN/SrTiO$_3$/Mn$_3$GaN AFMTJ (Supplementary Figs. S8c-f). While quantitively SOC reduces the ETMR ratio from 1.8×10$^4$% to 2.6×10$^3$%, qualitatively the ETMR effect remains huge driven by the nearly perfect spin polarization of Mn$_3$GaN.

## Discussion

In addition to the AFM $\Gamma_{5g}$ phase, there is another common noncollinear AFM configuration of antiperovskites known as the $\Gamma_{4g}$ phase[36,37]. The AFM $\Gamma_{4g}$ phase is obtained from $\Gamma_{5g}$ by rotating all magnetic moments about the [111] axis by 90°. The corresponding distribution of $s_{k_\parallel}$ in Mn$_3$GaN in the AFM $\Gamma_{4g}$ state mirrors this rotation (Fig. 5a, leftmost panel). This spin rotation, however, does not change the patterns of $s_{k_\parallel}$ and $p_{k_\parallel}$ (Figs. 5b,c, leftmost panels) which remain the same as those for the $\Gamma_{5g}$ phase. Therefore, the ETMR effect is also expected for Mn$_3$GaN/SrTiO$_3$/Mn$_3$GaN (001) AFMTJs with Mn$_3$GaN electrodes in the AFM $\Gamma_{4g}$ phase.

Apart from Mn$_3$GaN, other noncollinear AFM metals can be used as electrodes in AFMTJs. Figure 5 shows the calculated spin texture and effective spin polarization at the Fermi surface for different noncollinear antiferromagnets. It is evident that all of them exhibit a nearly perfect spin polarization in a substantial portion of the 2DBZ. The effective use of this high spin polarization in the TMR experiment requires an appropriate choice of a crystalline insulator to match its low-decay-rate evanescent states to the highly polarized states in the antiferromagnet. Especially promising in this regard is Mn$_3$NiN which has $p_{k_\parallel} \approx 100\%$ in a wide cross region around the 2DBZ center ($\bar{\Gamma}$ point) (Fig. 5c). This feature allows the use of the tunneling barriers, such as SrTiO$_3$ and MgO, in the respective AFMTJs: while the former supports efficient transmission along the $k_x = 0$ and $k_y = 0$ lines in the 2DBZ (Fig. 4b), the latter exhibits lowest decay rates around the $\bar{\Gamma}$ point[38]. On the other hand, using a conventional MgO barrier in AFMTJs with Mn$_3$Sn electrodes seems not the best choice, due to the nearly zero spin polarization around the $\bar{\Gamma}$ point in Mn$_3$Sn (Fig. 4c). This may be a reason why the TMR in Mn$_3$Sn/MgO/Mn$_3$Sn AFMTJs appeared to be relatively low[24].

The are a few important requirements for observing the predicted ETMR effect. AFMTJs must have identical AFM electrodes to provide matching between their spin states in the momentum space. High quality and crystallinity of AFMTJs are required for conservation of transverse momentum $k_\parallel$ in the process of tunneling. Defects in the barrier, such as oxygen vacancies, should be avoided because they can lead to diffuse scattering between different $k_\parallel$ thus suppressing ETMR (similar to TMR in conventional MTJs[39]). Also, measurements well below the Néel temperature are desirable, due to temperature-driven fluctuations of magnetic moments being detrimental for ETMR. Finally, measurements at a low bias voltage are required, since ETMR is expected to decrease with bias due to the mismatch of spin states at different energies.

We would like to emphasize that the momentum dependent spin polarization that is defined in this work is not the same as the transport spin polarization of non-collinear antiferromagnets in the diffusive transport regime[10,11]. While the latter represents the *net* spin polarization of charge carriers when transport occurs along certain crystallographic directions and is defined with respect to a global quantization axis, the former has, in general, different quantization axes for different transverse wave vectors. As a result, even in those cases when the net spin polarization is zero, the effect resulting from the momentum-dependent spin polarization does not generally vanish. Observing this requires, however, ballistic transport regime which can be realized in AFMTJs or in metallic structures whose dimensions are comparable or less that the mean free path of electrons, such as, for example, an atomically sharp domain wall recently observed in antiferromagnets [40]. Moreover, the concept of momentum dependent spin polarization is relevant not only to noncollinear antiferromagnets, but also to *all* materials hosting noncollinear spins in the momentum space. Particularly, nonmagnetic systems with broken space inversion symmetry are known to exhibit noncollinear spin textures due to spin-orbit coupling [41]. These systems may reveal a sizable $p_{k_\parallel}$ that can be efficiently utilized in ballistic spintronic devices.

Overall, the unique property of noncollinear antiferromagnets to sustain a nearly perfect transport spin polarization opens unprecedented opportunities for spintronics. Functionalizing this property in AMFTJs allows an efficient electric detection and control of the AFM Néel vector as a state variable. While the latter can be achieved via the predicted ETMR effect, the former is envisioned due to the potentially strong spin-transfer torques in these AFMTJs [42,43]. Therefore, noncollinear AFMTJs have potential to become a new standard for spintronics providing stronger magnetoresistive effects, few orders of magnitude faster switching speed, and much higher packing density than conventional MTJs.

Lastly, future generations of magnetic logic and memories are expected to operate at a sub 10 nm length scale that is comparable to or less than the mean free path of electrons. At these dimensions, ballistic transmission controls transport properties of materials, and thus the predicted nearly perfect spin polarization carried by electrons with noncollinear spins can be efficiently exploited in ballistic spintronic devices. Thus, our work opens a new paradigm for the investigation and application of spin-textured materials by employing momentum-dependent transport spin polarization.

## Methods

**Tight-binding model for a Kagome lattice.** We consider a 2D Kagome lattice with non-collinear magnetic moments, as shown in Figure 1a. Our tight-binding (TB) model assumes one orbital per atom with on-site spin splitting ($\Delta$), spin-independent first-nearest neighbor hoping ($t$), and magnetic moments oriented along the local axes given by the unit vectors $\hat{m}_j, j = 1, 2, 3$. The TB Hamiltonian in real space is given by

$$H = -t \sum_{\langle jj'\rangle\alpha} c_{j\alpha}^\dagger c_{j'\alpha} + \frac{\Delta}{2} \sum_{j\alpha} (\boldsymbol{\sigma}\cdot\hat{m}_j) c_{j\alpha}^\dagger c_{j\alpha}, \quad (3)$$

where $c_{j\alpha}^\dagger$ and $c_{j\alpha}$ are the creation and annihilation operators for site $j$ and spin $\alpha$, $\boldsymbol{\sigma}$ represents a vector of the Pauli matrices ($\sigma_x, \sigma_y, \sigma_z$), and summation $\langle jj'\rangle$ runs over the nearest-neighbor sites. Matrix elements of the TB Hamiltonian in the momentum space can then be written as follows:

$$H_{\alpha\alpha'}^{jj'}(\boldsymbol{k}) = \sum_i e^{i\boldsymbol{k}\cdot(\boldsymbol{R}_i+\boldsymbol{r}_{j'}-\boldsymbol{r}_j)} H_{\alpha\alpha'}^{jj'}, \quad (4)$$

where $H_{\alpha\alpha'}^{jj'}$ are matrix elements of Hamiltonian (3) in real space, $\boldsymbol{r}_j$ is the position vector of atom $j$ in the unit cell of the Kagome lattice, and $\boldsymbol{R}_i$ is the coordinate on the lattice cell $i$. With the three non-equivalent lattice sites, the TB matrix (4) has rank of 6. Its eigenvalues and eigenfunctions $\psi_{nk}$ are calculated numerically using the built-in function Eigensystem in Mathematica v13.1. The spin expectation values for each Bloch state $\psi_{nk}$ are calculated in the standard way: $\boldsymbol{s}_{nk} = \frac{\hbar}{2}\langle\psi_{nk}|\boldsymbol{\sigma}|\psi_{nk}\rangle$, where $\boldsymbol{k} = (k_x, k_x)$. To obtain the spin polarization of conduction channels at the Fermi energy $E_F$, we assume that $x$ is the transport direction. Then $\boldsymbol{p}_{k_y}$ for a given transverse wave vector $k_y$ is defined by

$$\boldsymbol{p}_{k_y}(E_F) = \frac{\sum_n \boldsymbol{s}_{nk_y}}{\sum_n |\boldsymbol{s}_{nk_y}|}, \quad (5)$$

where the spin expectation values are

$$\boldsymbol{s}_{nk_y}(E_F) = \frac{a}{\pi}\int \boldsymbol{s}_{nk}\,\delta(E_{nk}-E_F)dk_x. \quad (6)$$

**DFT calculations.** Calculations are performed within density functional theory (DFT) using a plane-wave pseudopotential method implemented in Quantum-ESPRESSO [44]. The ultrasoft pseudopotentials [45] and the generalized gradient approximation (GGA) [46] for exchange-correlation potential are employed in the calculations involving noncollinear magnetism. The plane-wave cut-off energy of 52 Ry and a $12\times12\times12$ $k$-point mesh in the irreducible Brillouin zone are used to achieve self-consistency in the electronic structure calculations for bulk antiferromagnets and a $12\times12\times1$ $k$-point mech in the calculations involving $Mn_3GaN/SrTiO_3$ (001) and $Mn_3GaN/SrTiO_3/Mn_3GaN/SrTiO_3$ (001) supercells used for AFMTJs with parallel and antiparallel Néel vector, respectively. The relaxed in-plane lattice parameters of $SrTiO_3$ $a = b = 3.94$ Å are assumed. Internal coordinates and the $c$-lattice constant of the supercell were relaxed until the force on each atom was less than 0.001 eV/Å. The resulting bond length between the Mn and O atoms at the $Mn_2N/TiO_2$ terminated $Mn_3GaN/SrTiO_3$ interface was found to be 2.087Å.

The spin expectation values for each Bloch state and spin polarizations of the conduction channels are obtained using a 50×50×50 $k$-point mesh. The layer-resolved spectral density (i.e. the layer- and $k_\parallel$-resolved density of states) of the relaxed $Mn_3GaN/SrTiO_3/Mn_3GaN$ (001) structure is calculated using supercell calculations involving periodic boundary conditions with gaussian broadening of 0.01 eV using 50×50 k-points. The effects of spin-orbit interaction on spin polarization of bulk $Mn_3GaN$ (001) and ETMR in $Mn_3GaN/SrTiO_3/Mn_3GaN$ (001) AFMTJs are evaluated using fully relativistic PAW pseudopotentials and discussed in Supplementary Information (Section H), while in the main text, these effects are neglected.

The quantum-transport calculations are performed using PWCOND code [47,48] implemented within Quantum ESPRESSO. In the calculations, the relaxed $Mn_3GaN/SrTiO_3/Mn_3GaN$ (001) structure is considered as the scattering region, ideally attached on both sides to semi-infinite Mn$_3$GaN leads. The $k_\parallel$-resolved transmission is obtained using 100×100 $k$-points in the 2D Brillouin zone. The total transmission as a function of energy is calculated using 50×50 $k$-points in the 2D Brillouin zone.

The decay rates of evanescent states in SrTiO$_3$ are obtained from its complex band structure calculated using PWCOND. An arbitrary wave vector consists of a component parallel to the interface, $\boldsymbol{k}_\parallel$, which is conserved during tunneling, and a component perpendicular to the interface, $k_z$. For each $\boldsymbol{k}_\parallel$, we calculate the dispersion relation $E = E(k_z)$, allowing complex $k_z = q + i\kappa$. The imaginary part $\kappa$ is the decay rate, so that the corresponding wave functions decay as $\sim e^{-\kappa z}$.

The figures are plotted using Matplotlib and FermiSurfer[49].

## Author contributions

G.G., D-F.S., and E.Y.T. conceived the idea. G.G. performed the primary DFT and quantum-transport calculations. M.E. carried out the TB modeling. Q.-Q.L. and D.-F.S. checked the validity of the DFT results. All authors contributed to interpreting and analyzing the results. E.Y.T. wrote the manuscript assisted by G.G., M. E., and D.-F.S. All authors contributed to the final version of the manuscript. E.Y.T. supervised this project.

## Acknowledgments

The authors thank Allan MacDonald and Kirill Belashchenko for useful discussions and Bimal Neupane for their help in the calculations. This work was supported by the National Science Foundation (grant No. DMR-2425567) (E.Y.T), Office of Naval Research (ONR grant N00014-20-1-2844) (G.G., E.Y.T.), and UNL's Grand Challenges catalyst award "Quantum Approaches Addressing Global Threats" (M.E., E.Y.T.). G.G. gratefully acknowledges support of a Junior Research Fellowship from Trinity College, Oxford, U.K. D.F.S. acknowledges support from the National Key R&D Program of China (grant No. 2021YFA1600200), the National Natural Science Foundation of China (grants Nos. 12274411, 12241405, and 52250418), the Basic Research Program of the Chinese Academy of Sciences (grant No. JZHKYPT-2021-08), and the CAS Project for Young Scientists in Basic Research (grant No. YSBR-084). Computations were performed at the University of Nebraska Holland Computing Center.

## Data availability

The data that supports the findings of this study are available from the corresponding authors.

## Competing interests

The authors declare no competing interests.

\* gautam.gurung@trinity.ox.ac.uk
† dfshao@issp.ac.cn
‡ tsymbal@unl.edu